\documentclass[longauth]{aaEC}

\usepackage{graphicx}
\usepackage{natbib}
\usepackage{scalerel}
\usepackage{comment}
\usepackage{amssymb}
\usepackage{latexsym}
\usepackage{longtable}
\usepackage{tikz}

\usepackage[table]{xcolor}

\usepackage{txfonts}
\usepackage[pdfencoding=auto,psdextra]{hyperref}
\hypersetup{
    colorlinks=true,
    linkcolor=blue,
    filecolor=magenta,      
    urlcolor=blue,
    citecolor=blue
}
\makeatletter
\renewcommand*\aa@pageof{, page \thepage{} of \pageref*{LastPage}}
\makeatother

\usepackage[utf8]{inputenc}

\usepackage[switch, modulo]{lineno}

\usepackage{euclid}

\graphicspath{{./}{figures/}}

\begin{document}

\title{\Euclid: Methodology for derivation of IPC-corrected conversion gain of nonlinear CMOS APS\thanks{This paper is published on behalf of the Euclid Consortium.}}

\newcommand{\orcid}[1]{} 
\author{J.~Le~Graet\orcid{0000-0001-6523-7971}\thanks{\email{legraet@cppm.in2p3.fr}}\inst{\ref{aff1}}
\and A.~Secroun\orcid{0000-0003-0505-3710}\inst{\ref{aff1}}
\and M.~Tourneur-Silvain\orcid{0009-0000-8031-1914}\inst{\ref{aff1}}
\and W.~Gillard\orcid{0000-0003-4744-9748}\inst{\ref{aff1}}
\and N.~Fourmanoit\orcid{0009-0005-6816-6925}\inst{\ref{aff1}}
\and S.~Escoffier\orcid{0000-0002-2847-7498}\inst{\ref{aff1}}
\and E.~Kajfasz\orcid{0000-0002-8575-405X}\inst{\ref{aff1}}
\and S.~Kermiche\orcid{0000-0002-0302-5735}\inst{\ref{aff1}}
\and B.~Kubik\orcid{0009-0006-5823-4880}\inst{\ref{aff2}}
\and J.~Zoubian\inst{\ref{aff1}}
\and S.~Andreon\orcid{0000-0002-2041-8784}\inst{\ref{aff3}}
\and M.~Baldi\orcid{0000-0003-4145-1943}\inst{\ref{aff4},\ref{aff5},\ref{aff6}}
\and S.~Bardelli\orcid{0000-0002-8900-0298}\inst{\ref{aff5}}
\and P.~Battaglia\orcid{0000-0002-7337-5909}\inst{\ref{aff5}}
\and D.~Bonino\orcid{0000-0002-3336-9977}\inst{\ref{aff7}}
\and E.~Branchini\orcid{0000-0002-0808-6908}\inst{\ref{aff8},\ref{aff9},\ref{aff3}}
\and M.~Brescia\orcid{0000-0001-9506-5680}\inst{\ref{aff10},\ref{aff11},\ref{aff12}}
\and J.~Brinchmann\orcid{0000-0003-4359-8797}\inst{\ref{aff13},\ref{aff14}}
\and A.~Caillat\inst{\ref{aff15}}
\and S.~Camera\orcid{0000-0003-3399-3574}\inst{\ref{aff16},\ref{aff17},\ref{aff7}}
\and V.~Capobianco\orcid{0000-0002-3309-7692}\inst{\ref{aff7}}
\and C.~Carbone\orcid{0000-0003-0125-3563}\inst{\ref{aff18}}
\and J.~Carretero\orcid{0000-0002-3130-0204}\inst{\ref{aff19},\ref{aff20}}
\and S.~Casas\orcid{0000-0002-4751-5138}\inst{\ref{aff21}}
\and M.~Castellano\orcid{0000-0001-9875-8263}\inst{\ref{aff22}}
\and G.~Castignani\orcid{0000-0001-6831-0687}\inst{\ref{aff5}}
\and S.~Cavuoti\orcid{0000-0002-3787-4196}\inst{\ref{aff11},\ref{aff12}}
\and A.~Cimatti\inst{\ref{aff23}}
\and C.~Colodro-Conde\inst{\ref{aff24}}
\and G.~Congedo\orcid{0000-0003-2508-0046}\inst{\ref{aff25}}
\and C.~J.~Conselice\orcid{0000-0003-1949-7638}\inst{\ref{aff26}}
\and L.~Conversi\orcid{0000-0002-6710-8476}\inst{\ref{aff27},\ref{aff28}}
\and Y.~Copin\orcid{0000-0002-5317-7518}\inst{\ref{aff2}}
\and F.~Courbin\orcid{0000-0003-0758-6510}\inst{\ref{aff29},\ref{aff30},\ref{aff31}}
\and H.~M.~Courtois\orcid{0000-0003-0509-1776}\inst{\ref{aff32}}
\and A.~Da~Silva\orcid{0000-0002-6385-1609}\inst{\ref{aff33},\ref{aff34}}
\and J.~Dinis\orcid{0000-0001-5075-1601}\inst{\ref{aff33},\ref{aff34}}
\and M.~Douspis\orcid{0000-0003-4203-3954}\inst{\ref{aff35}}
\and F.~Dubath\orcid{0000-0002-6533-2810}\inst{\ref{aff36}}
\and C.~A.~J.~Duncan\inst{\ref{aff26}}
\and X.~Dupac\inst{\ref{aff28}}
\and S.~Dusini\orcid{0000-0002-1128-0664}\inst{\ref{aff37}}
\and A.~Ealet\orcid{0000-0003-3070-014X}\inst{\ref{aff2}}
\and M.~Farina\orcid{0000-0002-3089-7846}\inst{\ref{aff38}}
\and S.~Farrens\orcid{0000-0002-9594-9387}\inst{\ref{aff39}}
\and F.~Faustini\orcid{0000-0001-6274-5145}\inst{\ref{aff40},\ref{aff22}}
\and S.~Ferriol\inst{\ref{aff2}}
\and M.~Frailis\orcid{0000-0002-7400-2135}\inst{\ref{aff41}}
\and E.~Franceschi\orcid{0000-0002-0585-6591}\inst{\ref{aff5}}
\and S.~Galeotta\orcid{0000-0002-3748-5115}\inst{\ref{aff41}}
\and B.~Gillis\orcid{0000-0002-4478-1270}\inst{\ref{aff25}}
\and C.~Giocoli\orcid{0000-0002-9590-7961}\inst{\ref{aff5},\ref{aff42}}
\and F.~Grupp\inst{\ref{aff43},\ref{aff44}}
\and S.~V.~H.~Haugan\orcid{0000-0001-9648-7260}\inst{\ref{aff45}}
\and W.~Holmes\inst{\ref{aff46}}
\and F.~Hormuth\inst{\ref{aff47}}
\and A.~Hornstrup\orcid{0000-0002-3363-0936}\inst{\ref{aff48},\ref{aff49}}
\and P.~Hudelot\inst{\ref{aff50}}
\and K.~Jahnke\orcid{0000-0003-3804-2137}\inst{\ref{aff51}}
\and M.~Jhabvala\inst{\ref{aff52}}
\and A.~Kiessling\orcid{0000-0002-2590-1273}\inst{\ref{aff46}}
\and M.~Kilbinger\orcid{0000-0001-9513-7138}\inst{\ref{aff39}}
\and R.~Kohley\inst{\ref{aff28}}
\and H.~Kurki-Suonio\orcid{0000-0002-4618-3063}\inst{\ref{aff53},\ref{aff54}}
\and P.~B.~Lilje\orcid{0000-0003-4324-7794}\inst{\ref{aff45}}
\and V.~Lindholm\orcid{0000-0003-2317-5471}\inst{\ref{aff53},\ref{aff54}}
\and I.~Lloro\inst{\ref{aff55}}
\and G.~Mainetti\orcid{0000-0003-2384-2377}\inst{\ref{aff56}}
\and D.~Maino\inst{\ref{aff57},\ref{aff18},\ref{aff58}}
\and E.~Maiorano\orcid{0000-0003-2593-4355}\inst{\ref{aff5}}
\and O.~Mansutti\orcid{0000-0001-5758-4658}\inst{\ref{aff41}}
\and O.~Marggraf\orcid{0000-0001-7242-3852}\inst{\ref{aff59}}
\and K.~Markovic\orcid{0000-0001-6764-073X}\inst{\ref{aff46}}
\and N.~Martinet\orcid{0000-0003-2786-7790}\inst{\ref{aff15}}
\and F.~Marulli\orcid{0000-0002-8850-0303}\inst{\ref{aff60},\ref{aff5},\ref{aff6}}
\and R.~Massey\orcid{0000-0002-6085-3780}\inst{\ref{aff61}}
\and E.~Medinaceli\orcid{0000-0002-4040-7783}\inst{\ref{aff5}}
\and S.~Mei\orcid{0000-0002-2849-559X}\inst{\ref{aff62}}
\and M.~Meneghetti\orcid{0000-0003-1225-7084}\inst{\ref{aff5},\ref{aff6}}
\and G.~Meylan\inst{\ref{aff29}}
\and M.~Moresco\orcid{0000-0002-7616-7136}\inst{\ref{aff60},\ref{aff5}}
\and L.~Moscardini\orcid{0000-0002-3473-6716}\inst{\ref{aff60},\ref{aff5},\ref{aff6}}
\and S.-M.~Niemi\inst{\ref{aff63}}
\and J.~W.~Nightingale\orcid{0000-0002-8987-7401}\inst{\ref{aff64}}
\and C.~Padilla\orcid{0000-0001-7951-0166}\inst{\ref{aff65}}
\and S.~Paltani\orcid{0000-0002-8108-9179}\inst{\ref{aff36}}
\and F.~Pasian\orcid{0000-0002-4869-3227}\inst{\ref{aff41}}
\and K.~Pedersen\inst{\ref{aff66}}
\and V.~Pettorino\inst{\ref{aff63}}
\and S.~Pires\orcid{0000-0002-0249-2104}\inst{\ref{aff39}}
\and G.~Polenta\orcid{0000-0003-4067-9196}\inst{\ref{aff40}}
\and M.~Poncet\inst{\ref{aff67}}
\and L.~A.~Popa\inst{\ref{aff68}}
\and F.~Raison\orcid{0000-0002-7819-6918}\inst{\ref{aff43}}
\and A.~Renzi\orcid{0000-0001-9856-1970}\inst{\ref{aff69},\ref{aff37}}
\and J.~Rhodes\orcid{0000-0002-4485-8549}\inst{\ref{aff46}}
\and G.~Riccio\inst{\ref{aff11}}
\and E.~Romelli\orcid{0000-0003-3069-9222}\inst{\ref{aff41}}
\and M.~Roncarelli\orcid{0000-0001-9587-7822}\inst{\ref{aff5}}
\and E.~Rossetti\orcid{0000-0003-0238-4047}\inst{\ref{aff4}}
\and R.~Saglia\orcid{0000-0003-0378-7032}\inst{\ref{aff44},\ref{aff43}}
\and D.~Sapone\orcid{0000-0001-7089-4503}\inst{\ref{aff70}}
\and B.~Sartoris\orcid{0000-0003-1337-5269}\inst{\ref{aff44},\ref{aff41}}
\and J.~A.~Schewtschenko\orcid{0000-0002-4913-6393}\inst{\ref{aff25}}
\and M.~Schirmer\orcid{0000-0003-2568-9994}\inst{\ref{aff51}}
\and G.~Seidel\orcid{0000-0003-2907-353X}\inst{\ref{aff51}}
\and M.~Seiffert\orcid{0000-0002-7536-9393}\inst{\ref{aff46}}
\and C.~Sirignano\orcid{0000-0002-0995-7146}\inst{\ref{aff69},\ref{aff37}}
\and G.~Sirri\orcid{0000-0003-2626-2853}\inst{\ref{aff6}}
\and L.~Stanco\orcid{0000-0002-9706-5104}\inst{\ref{aff37}}
\and J.~Steinwagner\orcid{0000-0001-7443-1047}\inst{\ref{aff43}}
\and P.~Tallada-Cresp\'{i}\orcid{0000-0002-1336-8328}\inst{\ref{aff19},\ref{aff20}}
\and D.~Tavagnacco\orcid{0000-0001-7475-9894}\inst{\ref{aff41}}
\and A.~N.~Taylor\inst{\ref{aff25}}
\and H.~I.~Teplitz\orcid{0000-0002-7064-5424}\inst{\ref{aff71}}
\and I.~Tereno\inst{\ref{aff33},\ref{aff72}}
\and R.~Toledo-Moreo\orcid{0000-0002-2997-4859}\inst{\ref{aff73}}
\and F.~Torradeflot\orcid{0000-0003-1160-1517}\inst{\ref{aff20},\ref{aff19}}
\and I.~Tutusaus\orcid{0000-0002-3199-0399}\inst{\ref{aff74}}
\and L.~Valenziano\orcid{0000-0002-1170-0104}\inst{\ref{aff5},\ref{aff75}}
\and T.~Vassallo\orcid{0000-0001-6512-6358}\inst{\ref{aff44},\ref{aff41}}
\and A.~Veropalumbo\orcid{0000-0003-2387-1194}\inst{\ref{aff3},\ref{aff9},\ref{aff76}}
\and Y.~Wang\orcid{0000-0002-4749-2984}\inst{\ref{aff71}}
\and J.~Weller\orcid{0000-0002-8282-2010}\inst{\ref{aff44},\ref{aff43}}}
										   
\institute{Aix-Marseille Universit\'e, CNRS/IN2P3, CPPM, Marseille, France\label{aff1}
\and
Universit\'e Claude Bernard Lyon 1, CNRS/IN2P3, IP2I Lyon, UMR 5822, Villeurbanne, F-69100, France\label{aff2}
\and
INAF-Osservatorio Astronomico di Brera, Via Brera 28, 20122 Milano, Italy\label{aff3}
\and
Dipartimento di Fisica e Astronomia, Universit\`a di Bologna, Via Gobetti 93/2, 40129 Bologna, Italy\label{aff4}
\and
INAF-Osservatorio di Astrofisica e Scienza dello Spazio di Bologna, Via Piero Gobetti 93/3, 40129 Bologna, Italy\label{aff5}
\and
INFN-Sezione di Bologna, Viale Berti Pichat 6/2, 40127 Bologna, Italy\label{aff6}
\and
INAF-Osservatorio Astrofisico di Torino, Via Osservatorio 20, 10025 Pino Torinese (TO), Italy\label{aff7}
\and
Dipartimento di Fisica, Universit\`a di Genova, Via Dodecaneso 33, 16146, Genova, Italy\label{aff8}
\and
INFN-Sezione di Genova, Via Dodecaneso 33, 16146, Genova, Italy\label{aff9}
\and
Department of Physics "E. Pancini", University Federico II, Via Cinthia 6, 80126, Napoli, Italy\label{aff10}
\and
INAF-Osservatorio Astronomico di Capodimonte, Via Moiariello 16, 80131 Napoli, Italy\label{aff11}
\and
INFN section of Naples, Via Cinthia 6, 80126, Napoli, Italy\label{aff12}
\and
Instituto de Astrof\'isica e Ci\^encias do Espa\c{c}o, Universidade do Porto, CAUP, Rua das Estrelas, PT4150-762 Porto, Portugal\label{aff13}
\and
Faculdade de Ci\^encias da Universidade do Porto, Rua do Campo de Alegre, 4150-007 Porto, Portugal\label{aff14}
\and
Aix-Marseille Universit\'e, CNRS, CNES, LAM, Marseille, France\label{aff15}
\and
Dipartimento di Fisica, Universit\`a degli Studi di Torino, Via P. Giuria 1, 10125 Torino, Italy\label{aff16}
\and
INFN-Sezione di Torino, Via P. Giuria 1, 10125 Torino, Italy\label{aff17}
\and
INAF-IASF Milano, Via Alfonso Corti 12, 20133 Milano, Italy\label{aff18}
\and
Centro de Investigaciones Energ\'eticas, Medioambientales y Tecnol\'ogicas (CIEMAT), Avenida Complutense 40, 28040 Madrid, Spain\label{aff19}
\and
Port d'Informaci\'{o} Cient\'{i}fica, Campus UAB, C. Albareda s/n, 08193 Bellaterra (Barcelona), Spain\label{aff20}
\and
Institute for Theoretical Particle Physics and Cosmology (TTK), RWTH Aachen University, 52056 Aachen, Germany\label{aff21}
\and
INAF-Osservatorio Astronomico di Roma, Via Frascati 33, 00078 Monteporzio Catone, Italy\label{aff22}
\and
Dipartimento di Fisica e Astronomia "Augusto Righi" - Alma Mater Studiorum Universit\`a di Bologna, Viale Berti Pichat 6/2, 40127 Bologna, Italy\label{aff23}
\and
Instituto de Astrof\'{\i}sica de Canarias, V\'{\i}a L\'actea, 38205 La Laguna, Tenerife, Spain\label{aff24}
\and
Institute for Astronomy, University of Edinburgh, Royal Observatory, Blackford Hill, Edinburgh EH9 3HJ, UK\label{aff25}
\and
Jodrell Bank Centre for Astrophysics, Department of Physics and Astronomy, University of Manchester, Oxford Road, Manchester M13 9PL, UK\label{aff26}
\and
European Space Agency/ESRIN, Largo Galileo Galilei 1, 00044 Frascati, Roma, Italy\label{aff27}
\and
ESAC/ESA, Camino Bajo del Castillo, s/n., Urb. Villafranca del Castillo, 28692 Villanueva de la Ca\~nada, Madrid, Spain\label{aff28}
\and
Institute of Physics, Laboratory of Astrophysics, Ecole Polytechnique F\'ed\'erale de Lausanne (EPFL), Observatoire de Sauverny, 1290 Versoix, Switzerland\label{aff29}
\and
Institut de Ci\`{e}ncies del Cosmos (ICCUB), Universitat de Barcelona (IEEC-UB), Mart\'{i} i Franqu\`{e}s 1, 08028 Barcelona, Spain\label{aff30}
\and
Instituci\'o Catalana de Recerca i Estudis Avan\c{c}ats (ICREA), Passeig de Llu\'{\i}s Companys 23, 08010 Barcelona, Spain\label{aff31}
\and
UCB Lyon 1, CNRS/IN2P3, IUF, IP2I Lyon, 4 rue Enrico Fermi, 69622 Villeurbanne, France\label{aff32}
\and
Departamento de F\'isica, Faculdade de Ci\^encias, Universidade de Lisboa, Edif\'icio C8, Campo Grande, PT1749-016 Lisboa, Portugal\label{aff33}
\and
Instituto de Astrof\'isica e Ci\^encias do Espa\c{c}o, Faculdade de Ci\^encias, Universidade de Lisboa, Campo Grande, 1749-016 Lisboa, Portugal\label{aff34}
\and
Universit\'e Paris-Saclay, CNRS, Institut d'astrophysique spatiale, 91405, Orsay, France\label{aff35}
\and
Department of Astronomy, University of Geneva, ch. d'Ecogia 16, 1290 Versoix, Switzerland\label{aff36}
\and
INFN-Padova, Via Marzolo 8, 35131 Padova, Italy\label{aff37}
\and
INAF-Istituto di Astrofisica e Planetologia Spaziali, via del Fosso del Cavaliere, 100, 00100 Roma, Italy\label{aff38}
\and
Universit\'e Paris-Saclay, Universit\'e Paris Cit\'e, CEA, CNRS, AIM, 91191, Gif-sur-Yvette, France\label{aff39}
\and
Space Science Data Center, Italian Space Agency, via del Politecnico snc, 00133 Roma, Italy\label{aff40}
\and
INAF-Osservatorio Astronomico di Trieste, Via G. B. Tiepolo 11, 34143 Trieste, Italy\label{aff41}
\and
Istituto Nazionale di Fisica Nucleare, Sezione di Bologna, Via Irnerio 46, 40126 Bologna, Italy\label{aff42}
\and
Max Planck Institute for Extraterrestrial Physics, Giessenbachstr. 1, 85748 Garching, Germany\label{aff43}
\and
Universit\"ats-Sternwarte M\"unchen, Fakult\"at f\"ur Physik, Ludwig-Maximilians-Universit\"at M\"unchen, Scheinerstrasse 1, 81679 M\"unchen, Germany\label{aff44}
\and
Institute of Theoretical Astrophysics, University of Oslo, P.O. Box 1029 Blindern, 0315 Oslo, Norway\label{aff45}
\and
Jet Propulsion Laboratory, California Institute of Technology, 4800 Oak Grove Drive, Pasadena, CA, 91109, USA\label{aff46}
\and
Felix Hormuth Engineering, Goethestr. 17, 69181 Leimen, Germany\label{aff47}
\and
Technical University of Denmark, Elektrovej 327, 2800 Kgs. Lyngby, Denmark\label{aff48}
\and
Cosmic Dawn Center (DAWN), Denmark\label{aff49}
\and
Institut d'Astrophysique de Paris, UMR 7095, CNRS, and Sorbonne Universit\'e, 98 bis boulevard Arago, 75014 Paris, France\label{aff50}
\and
Max-Planck-Institut f\"ur Astronomie, K\"onigstuhl 17, 69117 Heidelberg, Germany\label{aff51}
\and
NASA Goddard Space Flight Center, Greenbelt, MD 20771, USA\label{aff52}
\and
Department of Physics, P.O. Box 64, 00014 University of Helsinki, Finland\label{aff53}
\and
Helsinki Institute of Physics, Gustaf H{\"a}llstr{\"o}min katu 2, University of Helsinki, Helsinki, Finland\label{aff54}
\and
NOVA optical infrared instrumentation group at ASTRON, Oude Hoogeveensedijk 4, 7991PD, Dwingeloo, The Netherlands\label{aff55}
\and
Centre de Calcul de l'IN2P3/CNRS, 21 avenue Pierre de Coubertin 69627 Villeurbanne Cedex, France\label{aff56}
\and
Dipartimento di Fisica "Aldo Pontremoli", Universit\`a degli Studi di Milano, Via Celoria 16, 20133 Milano, Italy\label{aff57}
\and
INFN-Sezione di Milano, Via Celoria 16, 20133 Milano, Italy\label{aff58}
\and
Universit\"at Bonn, Argelander-Institut f\"ur Astronomie, Auf dem H\"ugel 71, 53121 Bonn, Germany\label{aff59}
\and
Dipartimento di Fisica e Astronomia "Augusto Righi" - Alma Mater Studiorum Universit\`a di Bologna, via Piero Gobetti 93/2, 40129 Bologna, Italy\label{aff60}
\and
Department of Physics, Centre for Extragalactic Astronomy, Durham University, South Road, Durham, DH1 3LE, UK\label{aff61}
\and
Universit\'e Paris Cit\'e, CNRS, Astroparticule et Cosmologie, 75013 Paris, France\label{aff62}
\and
European Space Agency/ESTEC, Keplerlaan 1, 2201 AZ Noordwijk, The Netherlands\label{aff63}
\and
School of Mathematics, Statistics and Physics, Newcastle University, Herschel Building, Newcastle-upon-Tyne, NE1 7RU, UK\label{aff64}
\and
Institut de F\'{i}sica d'Altes Energies (IFAE), The Barcelona Institute of Science and Technology, Campus UAB, 08193 Bellaterra (Barcelona), Spain\label{aff65}
\and
DARK, Niels Bohr Institute, University of Copenhagen, Jagtvej 155, 2200 Copenhagen, Denmark\label{aff66}
\and
Centre National d'Etudes Spatiales -- Centre spatial de Toulouse, 18 avenue Edouard Belin, 31401 Toulouse Cedex 9, France\label{aff67}
\and
Institute of Space Science, Str. Atomistilor, nr. 409 M\u{a}gurele, Ilfov, 077125, Romania\label{aff68}
\and
Dipartimento di Fisica e Astronomia "G. Galilei", Universit\`a di Padova, Via Marzolo 8, 35131 Padova, Italy\label{aff69}
\and
Departamento de F\'isica, FCFM, Universidad de Chile, Blanco Encalada 2008, Santiago, Chile\label{aff70}
\and
Infrared Processing and Analysis Center, California Institute of Technology, Pasadena, CA 91125, USA\label{aff71}
\and
Instituto de Astrof\'isica e Ci\^encias do Espa\c{c}o, Faculdade de Ci\^encias, Universidade de Lisboa, Tapada da Ajuda, 1349-018 Lisboa, Portugal\label{aff72}
\and
Universidad Polit\'ecnica de Cartagena, Departamento de Electr\'onica y Tecnolog\'ia de Computadoras,  Plaza del Hospital 1, 30202 Cartagena, Spain\label{aff73}
\and
Institut de Recherche en Astrophysique et Plan\'etologie (IRAP), Universit\'e de Toulouse, CNRS, UPS, CNES, 14 Av. Edouard Belin, 31400 Toulouse, France\label{aff74}
\and
INFN-Bologna, Via Irnerio 46, 40126 Bologna, Italy\label{aff75}
\and
Dipartimento di Fisica, Universit\`a degli studi di Genova, and INFN-Sezione di Genova, via Dodecaneso 33, 16146, Genova, Italy\label{aff76}}    



\abstract{
    We introduce a fast method to measure the conversion gain in Complementary Metal-Oxide-Semiconductors (CMOS) Active Pixel Sensors (APS), which accounts for nonlinearity and interpixel capacitance (IPC). The standard `mean-variance’ method is biased because it assumes pixel values depend linearly on signal, and existing methods to correct for nonlinearity are still introducing significant biases. While current IPC correction methods are prohibitively slow for a per-pixel application, our new method uses separate measurements of the IPC kernel to make an almost instantaneous calculation of gain. Validated using test data from a flight detector from the ESA \Euclid mission, the IPC correction recovers the results of slower methods within 0.1\% accuracy. Meanwhile the nonlinearity correction ensures an estimation of the gain that is independent of signal, correcting a bias of more than 2.5\% on gain estimation. 
}


   \titlerunning{Conversion gain of nonlinear infrared detectors}
   \authorrunning{Le Graët et al.}
   
   \maketitle


\section{Introduction}
\label{sec:intro}
Since the early 2000s, considerable efforts have been made to enhance the sensitivity of complementary metal-oxide-semiconductors (CMOS) imaging sensors \citep{Fossum-1997}. Currently, the industry is capable of producing large-format (exceeding a thousand pixels on each side) CMOS active pixel sensors (APS) that exhibit ultra-low noise with sensitivity ranging from UV to long-wavelength infrared (LWIR) wavelengths. Thanks to these advancements, CMOS APS are now highly suitable for low-light imaging across various wavelengths, making them ideal for both astronomical and cosmological observations.

Although CMOS APS with silicon-sensitive layers have been used in UV \citep{Greffe-2022} and visible light applications \citep{Soman-2015}, their most extensive development has occurred in the IR wavelength range. A significant number of CMOS APS for IR applications includes HgCdTe as the sensitive layer, taking advantage of its tunable bandgap \citep{Rogalski-2011}, which is adjusted by varying the composition of the alloy. Several companies have developed sensors with high performance, such as Raytheon with VIRGO sensor \citep{Bezawada-2004} or LYNRED with the ALFA sensor \citep{Gravrand-2022}. Nevertheless, the most used sensors in astronomical or cosmological missions are the HxRG series by Teledyne. These Teledyne detectors play a crucial role in a wide range of observatories. For instance, the H1RG model will be used in the ARIEL mission dedicated to exoplanet studies \citep{Pichon-2022} as well as the MAJIS instrument of the JUICE mission \citep{Cisneros-Gonzalez-2020}. Teledyne's detectors are deployed both in space for low light imaging missions \--- including the large H2RG-based focal plane array of \Euclid NISP spectro-photometer \citep{Mellier-2024, Jahnke-2024, secroun_2016}, the four H2RG-based instruments of JWST \citep{Rauscher-2014} and the H4RG-based spectro-imager of the \textit{Nancy Grace Roman} Space Telescope \citep{Mosby-2020}\---, and in ground-based observatories, including many of the Extremely Large Telescope (ELT) instruments \citep{Bezawada-2023}. Recently, CMOS APS based on avalanche photodiodes (APD) developed by the Leonardo company in collaboration with ESA and NASA \citep{Claveau-2022} have achieved performance comparable to classical HgCdTe sensors with large format arrays.

Regardless of the wavelength range, the design chosen to build them, or their scientific application, the performance of the sensors is generally evaluated through three fundamental parameters: quantum efficiency (QE), readout noise and dark current. These parameters are also used in the different pipelines to compute science data products. Assessing their absolute values requires expressing them in physical units (electrons) rather than arbitrary digital units (ADU) -- the standard unit of raw data. For that matter, the conversion gain, defined as the number of electrons represented by one ADU, is a crucial parameter of sensors and inaccuracies in gain estimation might bias that of the sensor's other parameters. A striking example is that of quantum efficiency (QE), evaluated as the ratio of the number of charges collected by the photodiode to the incident photons. The standard methods for measuring QE involve observing the change in ADU under a calibrated photon flux. Consequently, in order to obtain an absolute value of QE, the conversion gain must be accurately determined as any bias in gain measurement might propagate to the estimated QE.

Since conversion gain directly impacts the measurement of sensor performance, its measurement is subject to several technical requirements. In particular, with recent missions such as \Euclid and ARIEL pursuing ambitious scientific objectives, the required accuracy for conversion gain has increased significantly. Returning to the earlier example, the QE of the ARIEL AIRS detector must be measured with an accuracy better than 0.5\%, a requirement that directly constrains the uncertainty on the conversion gain. Similarly, the \Euclid mission requires photon signal estimates to be accurate to within 1\%, again increasing demands on gain precision. Finally these missions now specify a minimum fraction of operable pixels -- 95\% for \Euclid -- with operability defined by thresholds on performance such as QE, dark current, and readout noise. As a result, conversion gain must be measured at the pixel level. These combined requirements impose stringent constraints on conversion gain characterization.

One of the most challenging aspects of gain measurement is that gain functions as a black box. Within this black box, several physical processes may take place making the measurement sensitive to correlations with several other parameters, as well as environmental conditions. For instance, any correlation in the signal of closed pixels will bias the measurement of conversion gain \citep{moore_quantum_2006}. A known source of such spatial correlation is the electric cross-talk between neighboring pixels due to their proximty, referred to as interpixel capacitance (IPC). Moreover, as the gain have to be measure per pixel, this bias created by IPC on gain also need to be measure per pixel, making existing methods (detailed in Sect.~\ref{sec:limits}) hardly applicable. Similarly, the  temporal correlation arising from signal persistence (detailed in Sect.~\ref{sec:limits}) makes gain measurements particularly complex. Finally, while there is currently no published evidence to suggest a dependence on sensor temperature or photon wavelength, the signal dependence of gain measurement has been well-documented (detailed in Sect.~\ref{sec:nl_meanvar}) and remains an issue. As a result, using acquisitions with different signal levels will lead to different measurements of the conversion gain,  introducing systematic uncertainties that can easily exceed the required accuracy.
Given these challenges, there is a real need for a reliable method to measure an unbiased conversion gain at the pixel level, one that is decoupled from all other parameters.

The aim of this paper is to propose a new, easily applicable method for measuring a conversion gain that is decorrelated from interpixel capacitance (IPC) and independent of the signal level used during acquisition. In Sect.~\ref{sect:actual_gain}, we will detail the concept of conversion gain, explore classical measurement methods and discuss their limitations motivating the need for a new method of gain derivation. We will then, in Sect.~\ref{sect:ipc_corr}, propose and validate a new per pixel correction of IPC bias in gain measurement. Finally, in Sect.~\ref{sect:nl_corr},  we will introduce a new ``nonlinear'' mean-variance method, an original derivation of the relation between signal variance and mean, which allows to estimate a signal independent conversion gain. As this work is part of the \Euclid IR detectors characterization that was conducted at the Center for Particles Physics of Marseille (CPPM), the validation of both methods will be based on on-ground characterization data coming from CPPM campaigns (a detailed description of these data may be found in \citeauthor{secroun_2016} \citeyear{secroun_2016}).

\section{Conversion gain}
\label{sect:actual_gain}

\subsection{Conversion gain measurement}

Despite some variations in their design, CMOS APS broadly share a similar architectural framework, and their simplified design may be described as follows. Each pixel primarily comprises a photodiode (typically a reverse-biased p-n junction) for charge photo-generation. This photodiode, potentially followed by a multiplication region as in the case of APDs, is interfaced with the readout integrated circuit (ROIC) via an indium bump. The ROIC, which may vary in structure (e.g., source follower or capacitive transimpedance amplifier \citeauthor{Guellec-2017} \citeyear{Guellec-2017}), uses transistors to amplify and buffer the voltage signal. At the output of the ROIC, another buffer interfaces with external readout electronics, which generally include at least one analog-to-digital converter (ADC) channel to digitize the output voltage signal into arbitrary digital units (ADUs). Due to this architecture, the charge in each pixel can be read non-destructively. To reduce readout noise, each pixel is sampled repeatedly, generating a ``ramp'' of signal that can be processed using various techniques to estimate the flux 
\citep{Rauscher-2007}.
The conversion gain, expressed in electrons per ADU (\elec$\, {\rm ADU}^{-1}$), is the ratio of the number of electrons accumulated by the photodiode to the number of ADUs generated by the ADC. This gain is usually considered as the combination of three distinct processes, the charge to voltage conversion from electrons to Volts taking place in the photodiode, the amplification and buffering from Volts to Volts in the ROIC and the ADC conversion from Volts to ADU in the external electronics. Given its criticality in CMOS APS performance, considerable effort has been dedicated to accurately measuring conversion gain, leading to the development of various techniques. 
The capacitance comparison method \citep{finger_conversion_2005-1} enables precise measurement of conversion gain but necessitates adding a finely calibrated external capacitance. The Fe55 technique, commonly applied for CCDs \citep{Fraser-1994}, has been adapted for CMOS APS \citep{Fox_2009}. However, measuring conversion gain across every pixel with these techniques is particularly tedious, making them rarely used. In the end, the most prevalent method for measuring conversion gain involves flat-field acquisition, wherein all pixels are uniformly illuminated. Subsequently, employing one of the existing analytical methods, based on a statistical description of pixel response, allows for measuring the conversion gain for each pixel.

Recent studies \citep{Hendrickson_2023} have introduced methods that use sophisticated statistical descriptions of signal output to approximate measured distributions. These methods are specifically optimized for sub-electron readout noise detectors, such as APDs, making them less applicable to more common detectors with read noise ranges from a few electrons to tens of electrons. For these usual detectors, the ``gold standard'' method to measure gain using flat-fields was initially proposed by \citet{Mortara_1981} as the ``mean variance'' method and later refined by \citet{Janesick-2001a} into the well-known ``photon transfer curve'' (PTC). These methods are based on the derivation of the mean-variance equation, relating variance and mean of the output signal of a pixel. They presuppose that for a linear sensor, the output signal $S$ (ADU) of a pixel that has integrated a charge $Q$ (\elec) is given by
\begin{equation}
\label{eq:siglin}
    S = \dfrac{Q}{g}\;,
\end{equation}
where $g$ denotes the conversion gain in \elec$\, {\rm ADU}^{-1}$. Here, the readout noise does not appear as it is presumed to be Gaussian noise with a mean of zero and a standard deviation of $\sigma_r$. 

Assuming that $g$ is constant and that $Q$ does not depend on $g$, the variance $\sigma_S^2$ of the signal may be calculated by applying the error propagation formula to Eq.~\eqref{eq:siglin}. Moreover, the readout noise $\sigma_r$ should be added in quadrature in the variance equation,
\begin{equation}
    \sigma_S^2 = \left( \frac{\partial S}{\partial Q}\right)^2\left( \sigma_{Q}^2\right) + \left( \frac{\partial S}{\partial g}\right)^2 \left(\sigma_{g}^2 \right) + \sigma_r^2 \; .
\end{equation}
Finally, under the assumption of Poisson distributed integrated electrons ($\sigma_Q^2=Q$) as well as a negligible conversion gain variance $\sigma_g$, the total variance is expressed as
\begin{equation}
    \label{eq:meanvar}
    \sigma_S^2 =  \dfrac{S}{g} + \sigma_r^2 \; .
\end{equation}
According to Eq.~\eqref{eq:meanvar}, the conversion gain may be accurately determined by fitting linearly the mean-variance curve constructed from flat-field measurements. Typically, a mean-variance curve may be constructed from a series of $M$ similar flat-field ramps. Within these ramps, the signal variance and mean are estimated across the $M$ ramps for each pixel. As the ramps consist of measurements spaced by uniform integration times, the mean-variance curve will contain the same number of data points as there are measurements in a ramp. The gain is then derived as the inverse slope of the expected linear relationship, with the readout noise determined as its intercept.

\subsection{Uncertainty on gain measurement}
\label{sec:spatial}

Although some authors \citep{Beecken-1996} have attempted to estimate the uncertainty associated with conversion gain measurements using the mean-variance method, the practical difficulty in assessing the uncertainty of the variance used to construct the curve, followed by the fitting process, renders precise estimation challenging. Therefore, the conventional approach to uncertainty estimation involves repeating the measurement of a single pixel's gain and using the resulting mean as a gain estimator and the standard error of the sample mean, $\sigma_g/\sqrt{N}$, as the uncertainty estimator. Here, $N$ represents the number of gain measurements, and $\sigma_g$ denotes the standard deviation of these measurements. The challenge arises because $M$ ramps are needed to perform one gain measurement, and subsequently,  $N$ gain measurements require $N \times M$ ramps to estimate both the gain and the error. This requirement considerably increases the total number of acquisitions needed. For instance, aiming for an uncertainty of $1\,\%$, as for the \Euclid mission, and choosing to measure variance and mean across 50 ramps of 400 frames necessitates approximately 500 gain measurements, meaning to $25 \, 000$ ramps taking approximately 4000 hours (almost 6 months) in the case of \Euclid detectors, an impractical number. This estimation comes from a Monte Carlo simulation of an ideal linear sensor. 
A way to limit the amount of ramps needed is to combine spatial and temporal statistics by applying the ergodic hypothesis at small scales. This means making the assumption that the gain is the same for a box of $P$\texttimes$P$ pixels, named a superpixel, and that the values of signal in the $P$\texttimes$P$ pixels are uncorrelated. Then, rather than 50, only a pair of ramps acquired under the same conditions (flux, integration time, temperature, ...) are necessary to measure a per superpixel gain by estimating variance and mean spatially, across the $P$\texttimes$P$ pixels. At least two identical ramps are mandatory to eliminate fixed pattern noise \citep{Janesick-2001a}. 

\begin{figure}
   \centering
   \includegraphics[width=.8\columnwidth, angle=0]{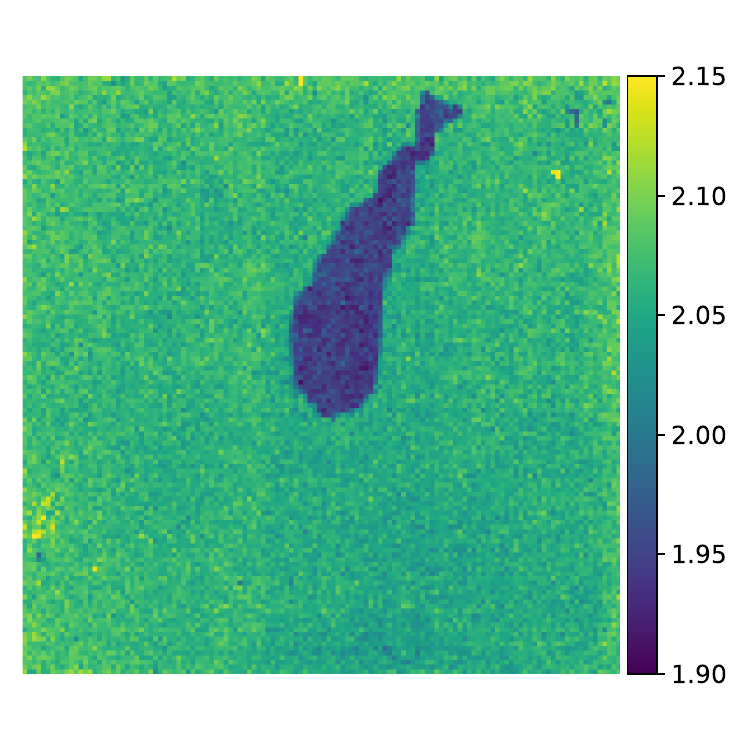}
   \caption{Map of the per-superpixel conversion gain (\elec${\rm ADU}^{-1}$) of a flight H2RG detector from the \Euclid mission, measured with the standard mean-variance method. The map reveals two distinct regions, a characteristic feature of many H2RG detectors, commonly caused by a lack of epoxy between the ROIC and the sensitive layer. Even within a single region, non-negligible gain variations remain.
   }
   \label{fig:bias_gain}
   \end{figure}

To demonstrate this methodology, it has been applied at CPPM to the ground characterization campaign data of one of the 16 H2RG flight detectors of the focal plane array of the near-infrared spectrophotometer of the \Euclid space mission (see \citeauthor{barbier_2018} \citeyear{barbier_2018} for information about the characterization campaign). The data used consists of pairs of flat-field ramps taken under fluxes between 20 and $1000\,{\rm photons}\,{\rm s}^{-1}$ at sensor temperature between $80\,{\rm K}$ and $90\,{\rm K}$ using a single Thorlabs 1600P LED illumination (1650~nm central wavelength at $300\,{\rm K}$). It was decided to combine data taken at different temperatures as our gain measurement showed no dependence on sensor temperature. Pairs were nevertheless made with ramps at the same temperature and illumination. Criteria for selecting pixels and ramps were implemented to ensure the method's assumptions (for detailed criteria, see \citeauthor{LeGraet-2022} \citeyear{LeGraet-2022}). For instance, ramps exhibiting signals below 70\% of the full well capacity ($\approx \, 130 \, {\rm k}$\elec for \Euclid)  were chosen to circumvent nonlinearity effects. The full well capacity is defined as the maximal charge a pixel can accumulate, beyond which additional incident photons no longer increase the signal level. For \Euclid it was estimated using a gain of $2\, e^{-}ADU^{-1}$. Superpixels were defined with dimensions of 16\texttimes16 pixels, and approximately 500 pairs of ramps were used. The 16\texttimes16 superpixel grid used here represents a standardized choice, however, the superpixel size and shape can be tailored to account for spatial heterogeneities in pixel behavior when necessary. The resulting per superpixel conversion gain map is depicted in Fig.\ref{fig:bias_gain}. This approach achieved an accuracy better than 1\% in gain estimation across all superpixels. However, it is known that spatial correlations among adjacent pixels exist \citep{finger_conversion_2005-1}, which could introduce biases in the estimation of the signal variance through spatial variance. Limitations of this spatial approach will be discussed in the following section.

\subsection{Limitations of the standard mean-variance method}
\label{sec:limits}

Three main parameters affecting the derivation of conversion gain are considered: nonlinearity, IPC, and persistence.
\paragraph{Nonlinearity}

\begin{figure}
   \centering
   \includegraphics[width=\columnwidth, angle=0]{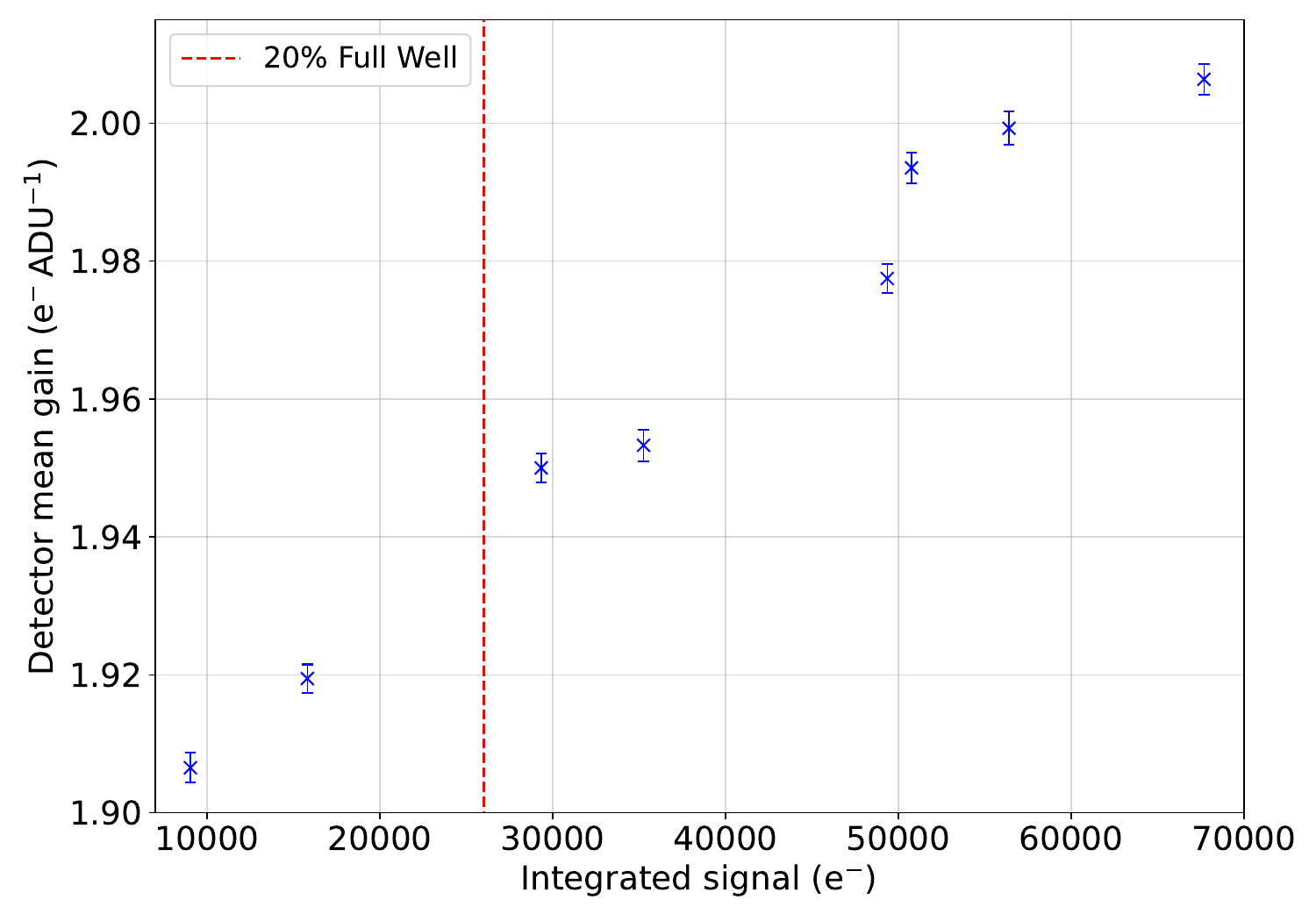}
   \caption{Conversion gain averaged across all pixels of a flight H2RG \Euclid detector measured at different integrated signals. The error bars include both statistical and systematic errors, the systematic errors are detailed in Sec.~\ref{sec:valid}. The signal dependence of the gain measurement is clearly visible, as the gain increases with the integrated signal.
   }
   \label{fig:gain1vsfluence}
   \end{figure}

To derive the mean-variance equation accurately, a linear relationship between the charge integrated by a pixel (in electrons) and its output signal (in ADU) is assumed. However, CMOS APS detectors typically exhibit nonlinear responses due to diode nonlinearity. This assumption, could lead to incorrect gain estimations. To assess the impact of integrated signal (number of electrons integrated by a pixel) on conversion gain accuracy, the 500 ramps used in Fig.\ref{fig:bias_gain} were divided by integrated signals. Subsequently, the gain for each integrated signal was calculated using the mean-variance method, averaging the results across the detector. The findings, presented in Fig.\ref{fig:gain1vsfluence}, reveal a significant dependence of measured gain on integrated signals, with variations exceeding 5\%.

The dependence of gain measurement on integrated signal, as determined by the mean-variance method, has been acknowledged for some time, resulting in various modifications to the Eq.~\eqref{eq:meanvar}. \citet{Pain-2003} introduced a gain definition that varies with integrated signal. This revised definition led to a new equation correlating the variance and mean of the signal, facilitating direct quantification of QE. However, this method is highly sensitive to its input parameters and does not allow for the direct measurement of conversion gain. It was suggested by
\citet{Janesick-2006} and \citet{Bezawada-2007} that both measuring conversion gain at integrated signals below 20\% of the full well capacity while ensuring predominance of photon shot noise could mitigate the influence of integrated signal on gain measurement. However, as illustrated in Fig.\ref{fig:gain1vsfluence} for the \Euclid detector, even below 20\% of full well (red dotted line), variations higher than 2\% can be seen. Given the precision requirements of modern missions, such variations must be accounted for, as they will introduce biases in the estimation of conversion gain.
In Sect.~\ref{sect:nl_corr} a new mean-variance equation that accounts for pixel response nonlinearity will be introduced to correct for the dependence of gain on integrated signal.

\paragraph{Inter Pixel Capacitance}

As explained in Sect.~\ref{sect:actual_gain}, the mean-variance method assumes that there are no correlations among the signals of the pixels in the detector. However, the phenomenon of Inter-Pixel Capacitance (IPC), detailed in Sect.~\ref{sect:ipc_corr}, resulting from electrostatic coupling between adjacent pixels, challenges this assumption by causing signal spread to neighboring pixels. IPC induces correlations between signals in adjacent pixels, which leads to an underestimation of signal variance and, consequently, an overestimation of the conversion gain. Different methods \citep{moore_quantum_2006, McCullough-2008} were proposed to adjust for IPC by incorporating spatial correlations into variance calculations. However, recent research \citep{LeGraet-2022} suggests that applying these methods could be highly complex due to the presence of other spatial correlations, mostly coming from persistence. Recently, \citet{hirata_2019} introduced a new model for output signal that incorporates pixel cross-talk (including IPC) and nonlinearity, allowing the calculation of temporal and spatial correlations in flat-field acquisitions. This model, when applied to flat-field correlations, allows for the estimation of several parameters, including conversion gain. Once again, the presence of unmodeled correlations biased the estimation of the different parameters. Furthermore, the need to fit the model to the flat-field correlations with many free parameters renders it very complex to use. In Sect.~\ref{sect:ipc_corr}, a simple and fast method will be presented to correct, on a per-pixel basis, the bias introduced by IPC in gain measurements obtained via the mean-variance method. This approach is universally applicable and only requires prior measurements of IPC.

\paragraph{Persistence}
One significant source of correlation that can bias previous methods is persistence. Persistence refers to a remnant signal coming from previous acquisitions, which results from charge trapping \citep{Smith-2008}. Consequently, during a flat-field acquisition, the apparent flux (flux measured between two frames) will not remain constant along the ramp, which will in the end affect the mean-variance curve.  Moreover, recent studies \citep{Ives2020, secroun_2018} highlight that spatial pattern exhibiting higher or lower persistence are common. This spatial feature of persistence leads to spatial correlations not accounted for in any previous models, resulting in biases when using spatial variance to estimate variance of the signal. While there is a need to properly model the effect of persistence in gain measurements, a preliminary approach is to mitigate it as much as possible. This is achieved by choosing ramps acquired when the sensor is in a steady state or nearly in a steady state. This steady state is defined as the time when the contribution of the persistence to the apparent flux becomes negligible compared to that from the illumination. For example, in the case of \Euclid, it has been demonstrated \citep{secroun_2018} that for ramps of 400 frames (1 frame every 1.41 s), after the first 100 frames, the apparent flux reaches a steady state. Also, for higher fluxes ($> 200\,{\rm photons}\,{\rm s}^{-1}$), it was observed that the steady state is reached when the integrated flux is greater than $10\,{\rm ke}^{-}$. Therefore, for all the ramps used in this study, either the first 100 frames or the frames before reaching an integrated flux of $10\,{\rm ke}^{-}$ were excluded.

\section{Correction of IPC impact on gain measurement}
\label{sect:ipc_corr}

\subsection{Inter pixel capacitance model}
In CMOS APS the very close proximity of pixels induces a parasitic capacitance between adjacent pixels, known as Inter Pixel Capacitance \citep{Moore-2003}. These parasitic capacitances cause electrical cross-talk between neighboring pixels, whereby a fraction of the charge generated in a pixel is detected in its neighbors. The classical model to describe the effect of IPC on the signal was well defined in \citet{moore_interpixel_2004}. In this paper, Moore et al. introduced $h$ as the 2D impulse response of a pixel, which describes where the charge generated in the central pixel is detected. Then the signal $S_{i,j \, {\rm meas}}$ detected in a pixel may be described as the convolution of the signal $S_{i,j \, {\rm true}}$ incoming onto the detector and the impulse response $h$ as seen in Eq.~\eqref{eq:ipc_sig}
\begin{equation}
\label{eq:ipc_sig}
    S_{i,j \, {\rm meas}} = S_{i,j \, {\rm true}} * h \; ,
\end{equation}
where $i$ and $j$ describe respectively the row and column positions of a pixel in the matrix.
In an ideal scenario without IPC, the impulse response, defined as a $3\times3$ kernel centered on the considered pixel, should be as
\begin{equation}
h =  \begin{bmatrix} 
    0 & 0 & 0 \\
    0 & 1 & 0 \\
    0 & 0 & 0
    \end{bmatrix} \; .
\end{equation}
Nevertheless, in the presence of IPC, the impulse response may adopt a more complex form. Various expressions of the impulse response with IPC exist, each either neglecting certain terms or assuming horizontal or diagonal symmetry. For example, Moore et al. assume that the diagonals are 0 and the cross coefficients are all equivalent, whereas \citet{Dudik-2012} assumes that all diagonals are equal to the squared values of the cross coefficients. With a view to unify the different expressions, the kernel may be written as
\begin{equation}
\label{eq:kernel_ipc}
h =  \begin{bmatrix} 
    \alpha_1 & \alpha_2 & \alpha_3 \\
    \alpha_4 & 1 - \sum \limits_{i=1}^8 \alpha_i & \alpha_5 \\
    \alpha_6 & \alpha_7 & \alpha_8 
    \end{bmatrix} \; .
\end{equation}
Here, a 3\texttimes3 matrix is used. However, if IPC also affects second neighbors, a 5\texttimes5 matrix should be considered. In addition to its effect on the signal, IPC also affects the spatial variance of the signal, thus biasing the estimation of variance of the signal used in the mean-variance method.

\subsection{New correction of IPC bias on gain measurement}
\label{sect:ipc_theo}
Because IPC smooths the charge distribution across pixels, it reduces the spatial variance of the signal. As a result, relying on spatial variance to estimate the signal variance could lead to biased measurements of conversion gain. Again, based on the definition of the IPC kernel, \citet{moore_interpixel_2004} develops a method to correct the bias induced by IPC on gain measurement. This method estimates the IPC effect through spatial correlations. However, as previously noted, other factors of spatial correlation like persistence may impede the application of this technique. Yet, the framework of Moore et al. remains applicable for defining IPC's impact on spatial variance. In that model, the variance estimator in the mean-variance method is given by
\begin{equation}
\label{eq:ipc_var}
    \mathrm{\hat{var}} = \sigma_S^2 \, \lVert h \rVert^2 \; .
\end{equation}
Here $\sigma_S^2$ represents the true variance of the signal in ADU and $\lVert h \rVert^2$ is the zero-lag autocorrelation of the impulse response. Generalizing the work of Moore et al. to Eq.~\eqref{eq:kernel_ipc}, the variance estimator becomes
\begin{equation}
    \mathrm{\hat{var}} = \sigma_S^2 \, \left[ \left(1 - \sum \limits_{i=1}^8 \alpha_i \right)^2 + \sum \limits_{i=1}^8 \alpha_i^2 \right] \; ,
\end{equation}
which may be written in a form similar to Eq.~\eqref{eq:ipc_var} as
\begin{equation}
\label{eq:ipc_coeff}
    \begin{split}
    & \mathrm{\hat{var}} = \sigma_S^2 \, k \; , \\
    & \mathrm{with} \; k = 1 - 2\sum \limits_{i=1}^8 \alpha_i + \left(\sum \limits_{i=1}^8 \alpha_i \right)^2 + \left(\sum \limits_{i=1}^8 \alpha_i^2 \right) \; .
    \end{split}
\end{equation}
A very similar equation can also be obtained with a 5\texttimes5 IPC kernel (see Annex~\ref{ann:ipc}). Equation~\eqref{eq:ipc_coeff} illustrates that the impact of IPC on the variance estimator manifests as a multiplier coefficient $k$, dependent solely on the IPC kernel. Notably, since IPC has been demonstrated to be independent of integrated signals \citep{Ives2020, LeGraet-2022}, a single coefficient per pixel suffices. Moreover, as IPC does not influence the mean signal estimated via spatial mean, this coefficient can be directly applied to the biased conversion gain. Thus, taking into account the IPC contribution to the signal variance and incorporating the corresponding variance into Eq.~\eqref{eq:meanvar}, an IPC-free conversion gain may be expressed as
\begin{equation}
\label{eq:ipc_corrg}
    g = k \, \hat{g} \; ,
\end{equation}
where $\hat{g}$ is the biased gain measured using the classical mean-variance method and $g$ the ``true'' gain, namely an IPC-free conversion gain. In the case of a per superpixel conversion gain, it will be necessary to derive a per superpixel multiplier coefficient, for instance, using the IPC averaged across the pixels included in the superpixel. Finally, the accurate derivation of this IPC-free gain hinges on the precise measurement of the IPC coefficients $\alpha_i$. Various methodologies for measuring these coefficients will be outlined in Sect.~\ref{sect:ipc_val}.

\subsection{Validation of the method}
\label{sect:ipc_val}
   
\begin{figure}
   \centering
   \includegraphics[width=1\columnwidth, angle=0]{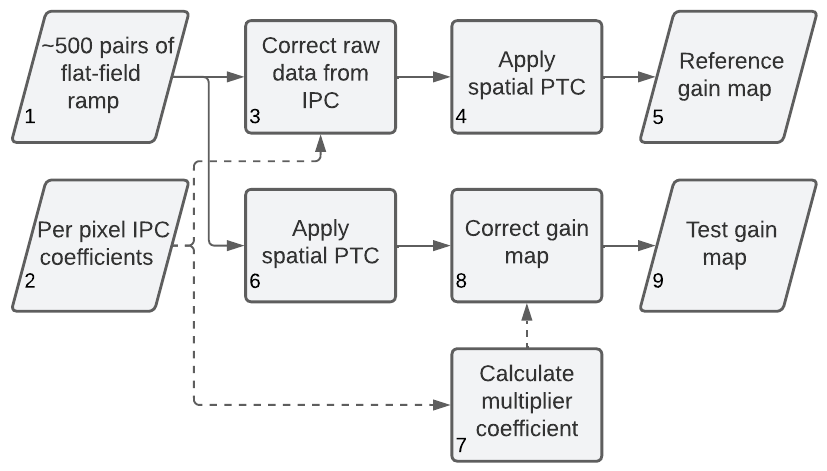}
   \caption{Pipeline used to validate the IPC-corrected gain method. The reference method, in the upper part of the diagram (boxes 3, 4), is used to produce a reference map, which is then compared to a map produced using the new method, in the lower part of the diagram (boxes 6, 7, 8).}
   \label{fig:ipc_valid}
   \end{figure}

To establish the effectiveness of the method proposed for correcting IPC effects on gain measurement, a proper validation process is essential. Figure~\ref{fig:ipc_valid} illustrates a block diagram of the validation pipeline, comparing the results of two distinct methods for obtaining an IPC-free gain. The new method, as outlined above, is depicted in the lower part of the diagram, while the reference method, previously validated in \citet{LeGraet-2022}, is shown in the upper part. 
In both cases, the IPC coefficients $\alpha_i$ must first be determined for each pixel (Fig.\ref{fig:ipc_valid} - box 2). Several methods, such as single-pixel illumination with $^{55}$Fe sources \citep{Fox_2009}, analysis of cosmic ray hits \citep{donlon_2017}, or uses of electrical charge generation in hot pixels \citep{Giardino_2012} have been employed for this purpose. However, only methods based on single pixel reset (SPR; see \citeauthor{seshadri_2008} \citeyear{seshadri_2008}) are efficient for measuring the IPC coefficients of each pixel of the sensor. When using these methods, one should be cautious to measure only IPC and not IPC combined with diffusion, as this model does not account for diffusion. Further studies on the impact of diffusion on spatial correlations are necessary to develop a model that accurately integrates these effects. For \Euclid's IR detectors, a modified SPR technique has been employed.

\renewcommand{\thefootnote}{$1$}

In the case of the reference method, going through the upper line of Fig.\ref{fig:ipc_valid}, these IPC coefficients may be used to directly correct IPC in the raw data in order to produce an IPC-free conversion gain map. Such a method is described in \citet{LeGraet-2022} and summarized here. Knowing the IPC kernel $k$ of each pixel (derived from previously measured IPC coefficients) enables the calculation of a corrective kernel, which, when convolved with the IPC kernel, results in the identity kernel. Then, the convolution of each pixel's raw data with its corresponding corrective kernel effectively corrects IPC effects (better than 1~\textperthousand). The latter convolution has thus been applied to all flat-field ramps used in the gain measurement (Fig.\ref{fig:ipc_valid} - box 3). Then implementing the mean-variance method with these ``IPC-free'' ramps (Fig.\ref{fig:ipc_valid} - box 4) leads to a conversion gain map that is unbiased by IPC, our reference gain map. One might question the direct application of this method to measure conversion gain. Actually, the need to correct the signal of each pixel in every frame of every ramp dramatically increases the computational time required. For instance, correcting the ramps in this study took as much time as acquiring them, which makes employing this method as a standard challenging.\footnote{Calculation were made using 22 Intel (R) Xeon(R) CPU E5-2640 v3 @ 2.60 GHz.} Nevertheless, applying it this one time gives us a useful reference conversion gain map (Fig.\ref{fig:ipc_valid} - box 5).

The new correction of IPC bias on gain measurements proposed here follows the second line of Fig.\ref{fig:ipc_valid}. First, the mean-variance method is applied to the same raw data set, producing a conversion gain map that initially includes IPC bias (Fig.\ref{fig:ipc_valid} - box 6). The per superpixel multiplier coefficients, derived from the $\alpha_i$ IPC coefficients as per Eq.~\eqref{eq:ipc_coeff}, are then used to determine an unbiased gain map according to Eq.~\eqref{eq:ipc_corrg} (Fig.\ref{fig:ipc_valid} - box 8), resulting in a test conversion gain map (Fig.\ref{fig:ipc_valid} - box 9). 

   \begin{figure}
   \centering
   \includegraphics[width=1\columnwidth, angle=0]{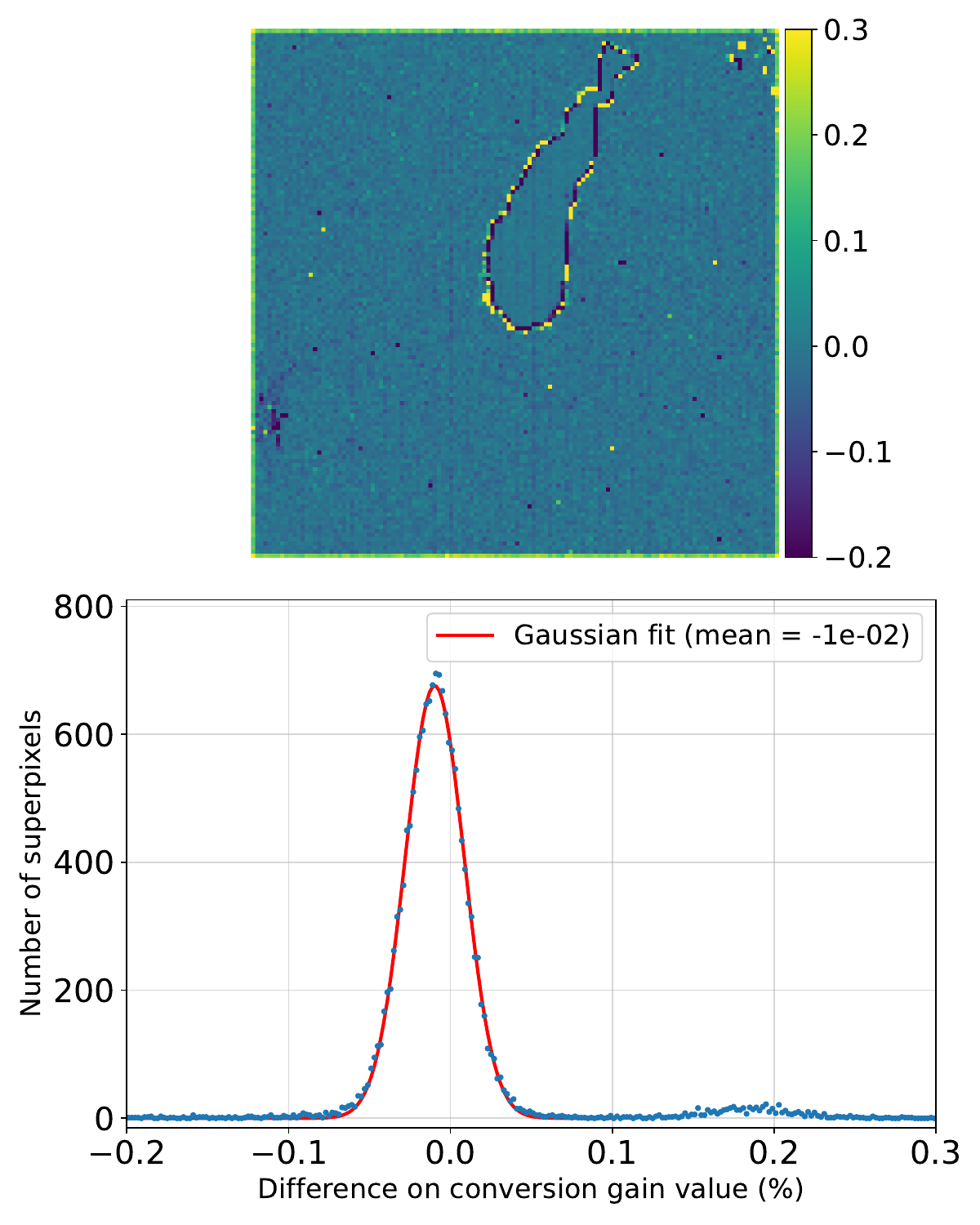}
   \caption{Map (top) and histogram (bottom) of the difference between the conversion gains of the reference map and the test map in percentage. The mean difference of 1~\textpertenthousand\ validates the new correction of the bias induced by IPC on gain measurement using mean-variance method.}
   \label{fig:ipc_corr}
   \end{figure}
   
\renewcommand{\thefootnote}{$2$}
To assess the efficacy of this new correction, the resulting test map is compared with the reference map. Fig.\ref{fig:ipc_corr} displays both the comparative map and its histogram that quantifies the relative difference (\%) between the two maps. The spatial mean of the difference was found to be less than 1~\textpertenthousand, significantly lower than the gain estimated error ($\approx \, 1\, \%$). This indicates that the two maps are equivalent, validating the use of multiplier coefficients to address IPC effects on conversion gain measurements. It may be noted that the histogram shows a minor peak at 0.19\%, corresponding to superpixels located at the boundary between reference and science pixels.\footnote{For H2RG detectors, the 2048\texttimes2048 matrix is divided into ``science'' pixels, the actual sensitive pixels, and ``reference'' pixels, frame of 4 external lines and columns of not sensitive pixels that surround the science pixels and that are usually used for noise correction.} Similarly, a marginally elevated difference is observed for superpixels along the boundary with the central region of the detector, an area known as the epoxy void \citep{secroun_2018}, which exhibits IPC lower than the rest of the detector. These two regions with larger differences arise from a limitation in the method employed to correct IPC in the raw data as detailed in \citet{LeGraet-2022}. Indeed, this method assumes that the IPC effect of the second neighbor on the first one is the same as the effect of the first neighbor on the central pixel. However, this difference is minimal, and even for affected superpixels, the discrepancy between the two methods is much smaller than the gain measurement error. Importantly, this bias does not originate from the new method itself. Therefore, this new approach is validated and may serve as a standard reference. It offers a considerably fast and light method (nearly instantaneous), necessitating only the prior measurement of IPC coefficients. Moreover, the method is compatible with superpixels of any size or shape -- including single pixels -- allowing adaptation to the specific spatial behavior of a given detector. In this study, a 16\texttimes16 pixel grid was chosen as a representative standard.

While this new method provides an easy-to-use tool for correcting the bias created by IPC on conversion gain, the bias coming from the signal dependence of gain measurement remains. In the following section, a new derivation of the mean-variance equation that corrects this dependence is presented.

\section{Correction of nonlinearity impact on gain measurement}
\label{sect:nl_corr}

Since conversion gain includes several components, as outlined in Sect.~\ref{sect:actual_gain}, resulting from the different stages of signal conversion and amplification, nonlinearity may emerge from each of these components. Firstly, the transistors used for voltage signal amplification and buffering are well-known sources of nonlinearity \citep{Hu-2010}. This form of nonlinearity is referred to as V/V nonlinearity. The second and more substantial source of nonlinearity is attributed to the charge-to-voltage conversion process, or V/Q nonlinearity, which is influenced by the varying pn-junction capacitance. Specifically, as photo-generated carriers are collected at the junction, the depletion region narrows, leading to a decrease of the voltage across the junction. Since the junction's capacitance is dependent on the applied voltage, it will decrease as charges accumulate.

Although they derive from distinct processes, V/V and V/Q nonlinearities are modeled through a common representation, typically, by a polynomial relation between the signal $S$ in ADU and the charge $Q$ in electrons. For instance, \citet{plazas_2017} and \citet{hirata_2019} used a second-order polynomial model of the pixel response as illustrated in Eq.~\eqref{eq:signal_nl},
\begin{equation}
    S = \dfrac{1}{g}(Q+\beta Q^2) \; .
    \label{eq:signal_nl}
\end{equation}
In this equation, the coefficient $\beta$, expressed in units of $e^{-1}$, encompasses both V/V and V/Q nonlinearities. These polynomial models are also implicitly used by different missions using CMOS APS \citep{Kubik-2014a, Canipe-2017} to correct the impact of nonlinearity on flux estimation. Precisely, the flux in ADUs per second is assumed to follow a polynomial model. Given that the photon flux and the conversion from photons to electrons (QE) are constant, the polynomial nature of flux in ADUs per second is presumed to arise from the conversion of electrons to ADU. For the correction of flux nonlinearity, depending on the mission, the order of the polynomial is often increased to achieve a better fit with the data. However, even though this polynomial flux correction has been commonly used, the mean-variance equation is still derived with a linear pixel response. In order to properly use mean-variance curve to estimate the conversion gain, this equation needs to be modified. In the following section, we will develop the mean-variance calculation incorporating nonlinear models. This updated nonlinear mean-variance equation will then be applied to measure the conversion gain

\subsection{Nonlinear mean variance method}
\label{sec:nl_meanvar}

Nonlinear models may be based on any order of polynomial. We will consider hereafter polynomials of order two and three.

\subsubsection{Second order polynomial model}
\label{sec:NL2}

First, let us use the Second order polynomial pixel response. Based on Eq.~\eqref{eq:signal_nl}, assuming that $g$, $Q$, and $\beta$ are independent (also meaning $g$ is constant), and adding the readout noise $\sigma_R$ as in Eq.~\eqref{eq:meanvar}, the error propagation formula may be written as
\begin{equation}
    \sigma_S^2 = \left( \frac{\partial S}{\partial Q}\right)^2\left( \sigma_{Q}^2\right) + \left( \frac{\partial S}{\partial g}\right)^2 \left(\sigma_{g}^2 \right) + \left( \frac{\partial S}{\partial \beta}\right)^2 \left(\sigma_{\beta}^2 \right) + \sigma_R^2 \; ,
\end{equation}
where $\sigma_S^2$ is the variance of the measured signal. Making the hypothesis that, for one pixel, the variances of $g$ and $\beta$ are negligible and knowing that the accumulated charge $Q$ follows a Poisson distribution, it follows
\begin{equation}
    \sigma_S^2 = \dfrac{1}{g^2} \left( 1 + 2\beta Q \right)^2 Q + \sigma_R^2 \; .
\end{equation}
Then, rearranging the terms, 
%
    \label{eq:pre_reform}
%
\begin{equation}
    \begin{split}
        \sigma_S^2 & =  \dfrac{1}{g^2}(Q + \beta Q^2) + \dfrac{3\beta}{g^2}Q^2( 1 + \beta Q + \beta^2\Bar{Q}^2) \\
        & + \dfrac{\beta^2}{g^2}Q^2(Q - 3\beta Q^2) + \sigma_R^2 \; ,
    \end{split}
    \label{eq:reform}
\end{equation}
and combining Eq.~\eqref{eq:signal_nl} squared with Eq.~\eqref{eq:reform} leads to
%
%
\begin{equation}
    \sigma_S^2 = \dfrac{1}{g} S + 3\beta S^2 + \dfrac{\beta^2}{g^2} Q^2( Q - 3\beta Q^2) + \sigma_R^2 \; .
    \label{eq:nl2_final}
\end{equation}
As mentioned previously, several missions apply polynomial fitting to their ramp data to correct for the impact of nonlinearity on flux measurements. This method also allows for an estimation of $\beta$ through the coefficient of the second order. According to recent estimations by the \Euclid characterization team $\beta$ is  evaluated on the order of $-5 \times 10^{-7}\,{\rm e}^{-1}$ \citep{Fourmanoit2-2021}. Similar estimates have been obtained by \citet{plazas_2017} and \citet{hirata_2019}. Hence, with this kind of value for $\beta$ and an integrated signal  lower than $70 \, {\rm k}$\elec, the ratio of the second term to the third term in Eq.~\eqref{eq:nl2_final} will be greater than 80. Therefore, the third term may be neglected. As a result, Eq.~\eqref{eq:nl2_final} simplifies to
\begin{equation}
    \sigma_S^2 \approx \dfrac{1}{g}S + 3\beta S^2 + \sigma_R^2 \; .
    \label{eq:nl_ptc}
\end{equation}
Equation~\eqref{eq:nl_ptc} is the basis of a new nonlinear mean-variance method that allows to derive the conversion gain $g$ in the presence of nonlinearities. In concrete terms, plotting a curve of the variance as a function of the mean of the signal for a pixel and fitting it with a second-order polynomial (rather than a first order) will allow the derivation of the conversion gain as the first order coefficient. Meanwhile, the nonlinearity coefficient $\beta$ may be obtained as that of second order. This new approach will be denoted below as the NL2 mean-variance.

\subsubsection{Third order polynomial model}
The same calculation applies using a Third order polynomial model for the pixel response as is given in Eq.~\eqref{eq:signal_nl3},
\begin{equation}
    \bar{S} = \dfrac{1}{g}(Q+\beta Q^2+\gamma Q^3) \; ,
    \label{eq:signal_nl3}
\end{equation}
with $\gamma$ the nonlinear third order coefficient in $e^{-2}$. Using the same assumptions as in Sect.~\ref{sec:NL2}, the variance of the signal may be written as
\begin{equation}
    \sigma_S^2 \approx \dfrac{1}{g}S + 3\beta S^2 + g\left( 5\gamma - 2\beta^2 \right) S^3 + \sigma_R^2 \; .
    \label{eq:nl3_ptc}
\end{equation}
Once again, fitting the mean-variance curve with a third-order polynomial will easily lead to the determination of gain from the first-order coefficient. This approach will be denoted as the NL3 mean-variance. In the following section, the standard mean-variance method along with the two nonlinear NL2 and NL3 adaptations introduced here will be applied to  and validated with \Euclid characterization data with a view to computing a gain corrected from nonlinearity.

\subsection{Validation of the nonlinear mean-variance method}

To assess and compare the gain values obtained from the three different mean-variance modeling approaches introduced here, namely linear, NL2, and NL3, two distinct strategies have been implemented. First the dependence of gain measurements on integrated signal using the three methods is examined and compared. This comparison helps to determine whether the new methods are compatible with the hypothesis of a gain  independent of integrated signal. Secondly, statistical tests will give hints to whether the new models lead to a significantly better assessment of the gain.

\subsubsection{Gain variations with integrated signal}
\label{sec:valid}
\begin{figure}
   \centering
   \includegraphics[width=1\columnwidth, angle=0]{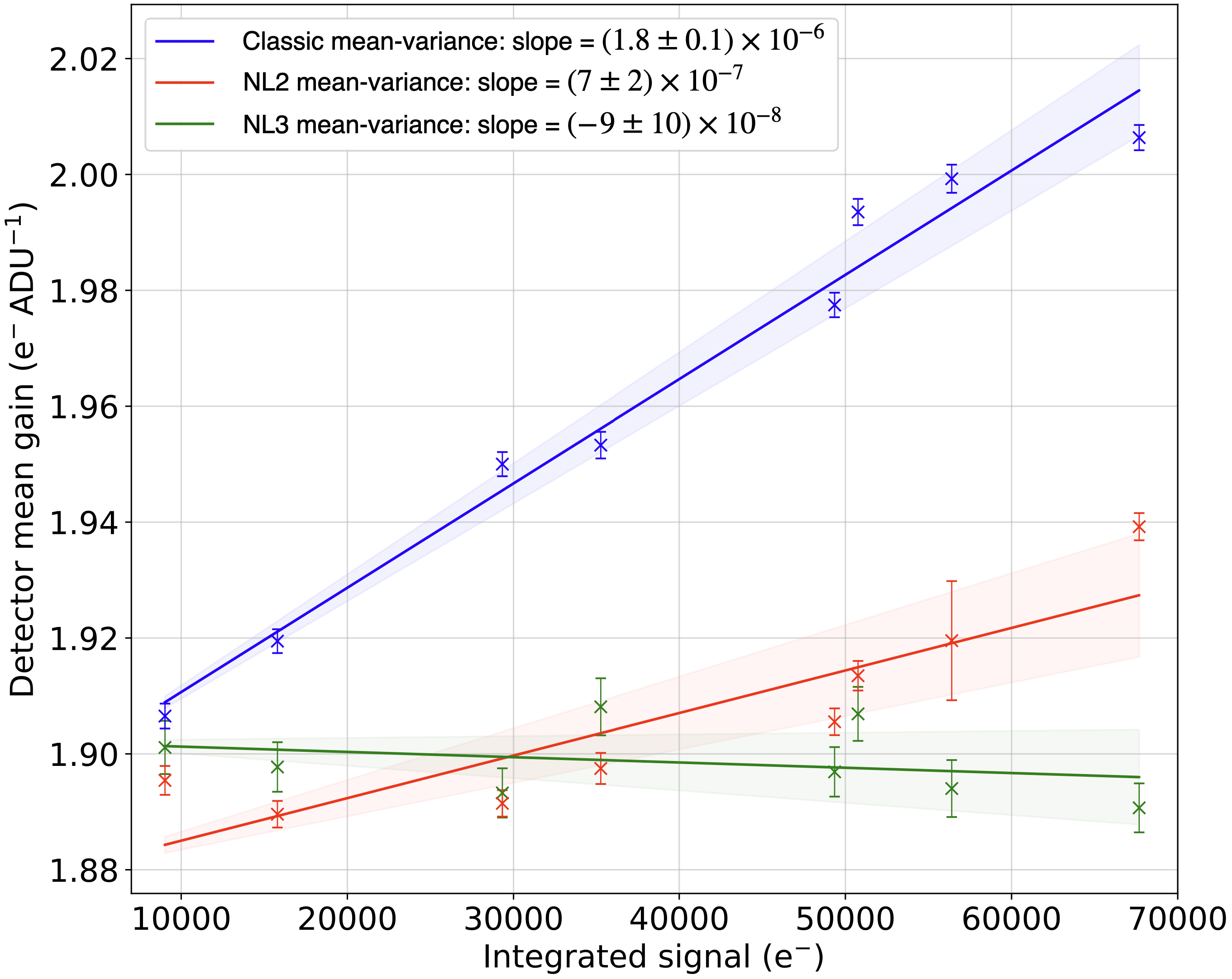}
   \caption{Conversion gain averaged across all pixels of a flight H2RG \Euclid detector measured at different integrated signals with linear, NL2 and NL3 mean-variance method. For each method, a linear regression and its 1-sigma confidence interval are plotted. Only the NL3 mean-variance method is consistent with the hypothesis that the gain remains constant with integrated signal. Mean gain measurement is $2.5 \, \%$ lower when using NL3 mean-variance instead of linear one. }
   \label{fig:gain3vsfluence}
   \end{figure}

To evaluate the efficiency of the three methods, they have been applied to the same data set referenced in Sect.~\ref{sect:ipc_corr}. As mentioned earlier, this data set consists of flat-field pairs of ramps acquired under various fluxes with constant integration time. By dividing the data set into smaller subsets with the same integrated signal and measuring the conversion gain for each subset, the influence of integrated signal on gain measurement may be investigated. In Fig.\ref{fig:gain3vsfluence}, the three mean-variance methods are applied to these subsets and the mean detector gain thus derived plotted against integrated signal in electrons (estimated from LED calibration). Gain values can be averaged across the detector thanks to the verified uniformity of the pixel response nonlinearity. As the gain is averaged across the whole detector, the statistical error is very low and the error bars are dominated by systematic error. This systematic error arises from flux-dependent effects, such as persistence, leading to variations in gain measurement for pairs of ramps acquired under different fluxes even with identical integrated signal. 

As previously seen in Fig.\ref{fig:gain1vsfluence}, the gain measured with the classical linear mean-variance equation (drawn in blue in Fig.\ref{fig:gain3vsfluence}) clearly increases with integrated signal, which is incompatible with the hypothesis that gain is independent of integrated signal. This behavior may be attributed, at least in part, to some nonlinearity related to the reduction of the junction capacitance. Regarding the two nonlinear mean-variance methods, NL2 and NL3, the detector's mean coefficient $\beta$ has been measured at, respectively, $(-4.2\pm0.1) \times 10^{-7}\,{\rm e}^{-1}$ and $(-5.6 \pm 0.6) \times 10^{-7}\,{\rm e}^{-1}$. Besides validating the assumption to derive Eq.~\eqref{eq:nl2_final} and Eq.~\eqref{eq:nl3_ptc}, both values are consistent with the $\beta$ estimation of the \Euclid characterization team. Using the NL2 mean-variance model (drawn in red in Fig.\ref{fig:gain3vsfluence}), it may be seen that for integrated signal levels lower than $50\,{\rm ke}^{-}$, the measured gain aligns with the expectation of a constant gain. Yet, at higher integrated signal levels, the gain starts to increase with integrated signal once more, indicating that a second-order polynomial is not sufficient for fully capturing the nonlinearity of the mean-variance curve.  In contrast, the gain measured using the NL3 mean-variance model (drawn in green in Fig.\ref{fig:gain3vsfluence}) remains consistent with a constant gain, within uncertainties, across all integrated signal levels between 10 and $70\,{\rm ke}^{-}$ ($\approx 60\,\%$ of the full well capacity). These findings strongly indicate that a third-order polynomial is best suited for accurately modeling the mean-variance curve and consequently deriving a truly constant gain value. When using the NL3 mean-variance model instead of the linear one, we correct a mean bias of about $2.5 \, \%$ on the gain measurement.

\subsubsection{Complementary validation with statistical tests}
\label{sec:stat2}
To conclusively validate the NL3 mean-variance method, a strict statistical analysis is required. This analysis will be conducted using a partial F-test. This test is designed to assess the significance of adding or removing variables from a regression model. First, it is necessary to define the null hypothesis to be tested. In this context, the null hypothesis assumes that the model with fewer variables fits the mean-variance curve as effectively as the models with additional variables. Should the null hypothesis not be rejected, it implies that the model with more variables does not offer a significantly improved fit.

For the evaluation of the null hypothesis, the partial F-test relies on comparing the sum of square residuals (SSR) between the two considered model fits. Hereafter, ${\rm SSR}_1$ will represent the SSR for the models with fewer variables, while ${\rm SSR}_2$ will denote the SSR for the alternative model. The respective numbers of parameters for these models are denoted as $p_{1}$ and $p_{2}$ and $n$ is the number of points used to fit the model (the number of frames in our study). Under the null hypothesis, the ratio defined in Eq.~\eqref{eq:ftest} follows a Fisher distribution with degrees of freedom $(p_{2}-p_{1},n-p_{2})$. For each of the mean curve variance used to measure the gain, the SSR from the fits with the two models under comparison, using Eq.~\eqref{eq:ftest},
\begin{equation}
    \label{eq:ftest}
    f_{{\rm value}} = \frac{\left(\frac{{{\rm SSR}}_{1}-{{\rm SSR}}_{2}}{p_{2}-p_{1}}\right)}{\left(\frac{{{\rm SSR}}_{2}}{n-p_{2}}\right)} \; ,
\end{equation}
allows the calculation of the $f_{\rm value}$. To refute the null hypothesis, the cumulative distribution function of a Fisher ($\mathcal{F}$) distribution with degrees of freedom $(p_{2}-p_{1},n-p_{2})$, evaluated at the $f_{\rm value}$, is computed. If the probability $\operatorname{P}(\mathcal{F} \leq f_{\rm value}) > 1 - \alpha$ for a given significance level $\alpha$, then the null hypothesis can be dismissed and the test is a success. A successful test leads to the conclusion that the model with additional variables provides a better fit than the one with fewer variables. A typical significance level is set to 0.05 (probability of $95\,\%$), but when multiple comparisons are performed, the probability of incorrectly rejecting the null hypothesis increases. In this case, since an F-test will be performed for each mean-variance curve used to measure a gain value, the significance level needs to be adjusted accordingly. To address this issue, a Bonferroni correction is applied to the significance level, given by $1 - \dfrac{\alpha}{M}$, where $M$ is the number of comparisons.

   \begin{figure}
   \centering
   \includegraphics[width=1\columnwidth, angle=0]{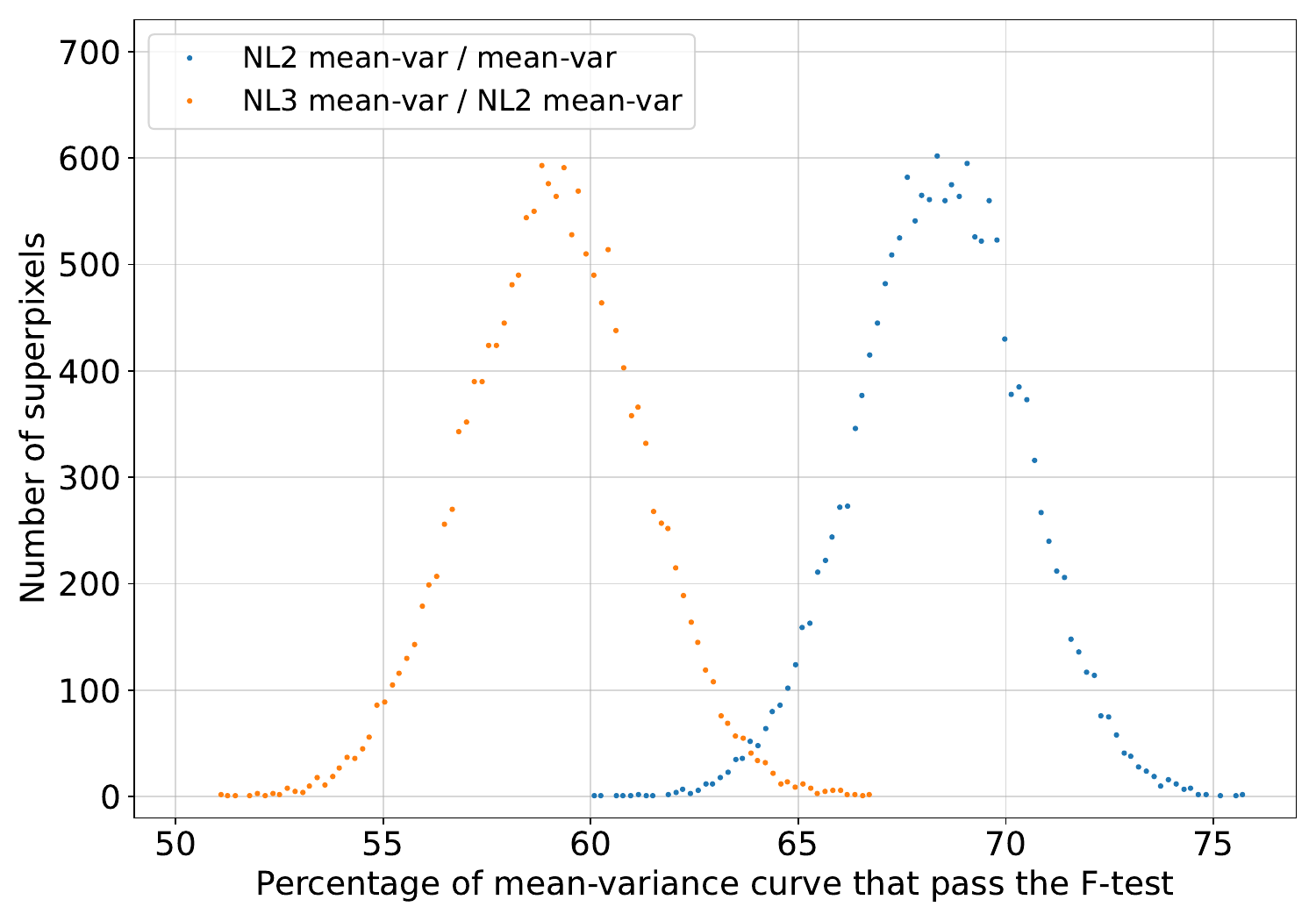}
   \caption{Distribution across the superpixels of the percentage of mean-variance curves that pass the F-test for two comparisons: NL2 mean-var vs. standard mean-var and NL3 mean-var vs. NL2 mean-var. For a superpixel, if more than $50\,\%$ of the mean-variance curve passes the test, the new model is validated.}
   \label{fig:ftest}
   \end{figure}

Two partial F-tests have been conducted: one to assess whether the NL2 mean-variance model outperforms the linear model, and another to determine if the NL3 mean-variance model provides a better fit than the NL2 model. As these evaluations are performed for all the mean-variance curves, the outcome will be for each curve, whether or not the F-test is a success. Precisely the percentage of mean-variance curves that pass the test is computed for each superpixel and their distribution is represented in Fig.\ref{fig:ftest}. If more than half of the mean-variance curves pass the test, the model with more variables is validated.

As observed, for all superpixels, more than $60\,\%$ of the mean-variance curves are more accurately represented by the NL2 model than by the linear model. Moreover, for more than $50\,\%$ of the mean-variance curves across all superpixels, the NL3 model successfully passes the test. This confirms that the NL3 mean-variance model is preferable for measuring the conversion gain of the \Euclid IR detectors. Investigations have begun into the use  of fourth-order model, however, deriving the variance formulas for such models becomes extremely complex, and for the \Euclid detector, the fourth order is dominated by noise. Therefore, for now, the NL3 mean-variance model remains the best suited method for measuring a conversion gain independent of integrated signal.

\section{Conclusions}
\label{sect:conclusion}
This study presents an innovative method for accurately measuring the conversion gain in CMOS Active Pixel Sensors, particularly focusing on correcting the biases created by both IPC and nonlinearity. The derivation of a new mean-variance equation, based on a polynomial description of the pixel response, allows the definition of a nonlinear mean-variance method that takes into account detector nonlinearity.  Furthermore, to correct the bias created by IPC on gain measurements, the method only uses a multiplier coefficient calculated from per pixel IPC coefficients, making it both faster and more efficient than typical methods. Applied to on-ground characterization data from a \Euclid infrared detector, this method underwent rigorous validation, demonstrating its effectiveness in correcting IPC bias and measuring a gain independent of integrated signal. As it allows a per-superpixel measurements of conversion gain decorrelated from IPC and nonlinearity and requiring only the knowledge of IPC coefficients, this new method, applicable to any CMOS APS could be used as a new reference.

\begin{acknowledgements}
This work was developed within the frame of a CNES-CNRS funded Phd thesis. \AckEC
\end{acknowledgements}

\bibliography{sample631}{}

\appendix

\section{Gain corrective coefficient for 5x5 IPC kernel}
\label{ann:ipc}
If the IPC between a pixel and his second neighbors is not negligible, the impulse response of a pixel needs to be a 5\texttimes5 kernel as
\begin{equation}
\label{eq:bigkernel_ipc}
h =  \begin{bmatrix} 
    \beta_1 & \beta_2 & \beta_3 & \beta_4 & \beta_5 \\
    \beta_6 & \alpha_1 & \alpha_2 & \alpha_3 & \beta_7 \\
    \beta_8 & \alpha_4 & 1 - \sum \limits_{i=1}^8 \alpha_i & \alpha_5 & \beta_9 \\
    \beta_{10} & \alpha_6 & \alpha_7 & \alpha_8 & \beta_{11} \\
    \beta_{12} & \beta_{13} & \beta_{14} & \beta_{15} & \beta_{16} 
    \end{bmatrix} 
\end{equation} 
Then, the corrective coefficient of Eq.~\eqref{eq:ipc_coeff} becomes :
\begin{align}
\label{eq:bigipc_coeff}
    k = &  \left(1 - \sum \limits_i \alpha_i - \sum \limits_j \beta_j  \right)^2 + \sum \limits_i \alpha_i^2 + \sum \limits_j \beta_j^2 \\
      = &  \, 1 - 2 \left( \sum \limits_i \alpha_i + \sum \limits_j \beta_j \right)  + 2 \sum \limits_i \alpha_i \sum \limits_j \beta_j \\
      & + \left(\sum \limits_i \alpha_i \right)^2 + \left(\sum \limits_j \beta_j^2 \right)^2 + \sum \limits_i \alpha_i^2 + \sum \limits_j \beta_j^2
\end{align}


\label{LastPage}
\end{document}